\documentclass[a4paper,11pt]{article}
\pdfoutput=1 % if your are submitting a pdflatex (i.e. if you have
\usepackage{jcappub}
\usepackage[T1]{fontenc}

\newcommand{\beq}{\begin{equation}}
\newcommand{\eeq}{\end{equation}}
\newcommand{\beqa}{\begin{eqnarray}}
\newcommand{\eeqa}{\end{eqnarray}}

\newcommand{\Tb}{{\overline T}}
\newcommand{\Neff}{N_{\rm eff}}

\usepackage[dvipsnames]{xcolor}

\title{Photon to axion conversion during Big Bang Nucleosynthesis}

\author[a,b]{A.J.~Cuesta,}
\author[b,1]{J.I.~Illana,\note{Corresponding author.}}
\author[b]{M.~Masip}

\affiliation[a]{Departamento de F{\'\i}sica, Universidad de C\'ordoba, \\ E-14071 C\'ordoba, Spain}
\affiliation[b]{CAFPE and Departamento de F{\'\i}sica Te\'orica y del Cosmos,
Universidad de Granada, \\ E-18071 Granada, Spain}

\emailAdd{ajcuesta@uco.es}
\emailAdd{jillana@ugr.es}
\emailAdd{masip@ugr.es}

\abstract{We investigate how the resonant conversion at a temperature $\Tb=25$--$65$~keV of a fraction of the CMB photons into an axion-like majoron affects BBN. The scenario, that assumes the presence of a primordial magnetic field and the subsequent decay of the majorons into neutrinos at $T\approx 1$ eV, has been proposed to solve the $H_0$ tension. We find two main effects. First, since we lose photons to majorons at $\Tb$, the baryon to photon ratio is smaller at the beginning of BBN ($T>\Tb$) than during decoupling and structure formation ($T\ll \Tb$). This relaxes the $2\sigma$ mismatch between the observed deuterium abundance and the one predicted by the standard $\Lambda$CDM model. Second, since the conversion implies a sudden drop in the temperature of the CMB during the final phase of BBN, it interrupts the synthesis of lithium and beryllium and reduces their final abundance, possibly alleviating the lithium problem.}

\begin{document}
\maketitle
\flushbottom

\section{Introduction\label{sec1}}

BBN probes our cosmological model at temperatures between 10 keV and 1 MeV, which are much higher than the ones probed by CMB and expansion-rate observables. The physics involved in nucleosynthesis has been 
established in lab experiments,   providing a very consistent framework to understand the early universe as well. During the past decade, however, cosmology has entered a precision era able to test the framework to a new level \cite{Steigman:2007xt}. In particular, the standard 
$\Lambda$CDM model depends on just a few free parameters, with two of them affecting critically both BBN and the observables at lower temperatures.

The first one is the baryon to photon ratio or $\eta_{10}$,
\beq
\eta_{10}= 10^{10}\,{n_b\over n_\gamma}\,. 
\eeq
A larger number of baryons at BBN implies a weaker deuterium bottleneck and, generically, a larger ratio of primordial products
($^4$He, $^7$Li) to reactives (D, $^3$H) \cite{Schramm:1997vs}. At lower temperatures, during the formation of large scale structures (LSS) a larger number of baryons leads to more efficient acoustic oscillations \cite{Eisenstein:1997ik}\footnote{See \href{http://background.uchicago.edu/~whu/animbut/anim1.html}{http://background.uchicago.edu/$\sim$whu/animbut/anim1.html}.} and thus to more pronounced anisotropies in the CMB \cite{Hu:1995kot}.

The second parameter is the total energy density $\rho_\nu$ carried by dark radiation (neutrinos and any other light relics) relative to photons. This is usually parametrized in terms of the effective number of neutrino species $N_{\rm eff}$,
\beq
\rho_\nu =  {7\over 8}\,{\pi^2\over 15} \,T_\nu^{4} \, N_{\rm eff}
\eeq
with $T_\nu=\left(\frac{4}{11}\right)^{\!1/3}T$ \cite{Lesgourgues:2013sjj}. A larger value of $\rho_\nu$ tends to increase the decoupling temperature and thus the number of neutrons relative to protons, implying a larger abundance of $^4$He \cite{Cyburt:2015mya}. At lower temperatures, it delays the time of matter-radiation equality and suppresses the power spectrum at small scales, as neutrinos free-stream during LSS formation \cite{Bond:1980ha}. On one hand, a larger $\rho_\nu$ and an increased expansion rate at recombination affects the position and amplitude of the peaks in the CMB spectrum, observables that are consistent with the standard value $N_{\rm eff}=3.043$ \cite{Akita:2020szl,Froustey:2020mcq,Bennett:2020zkv,Cielo:2023bqp}. On the other hand, it implies a larger expansion rate $H_0$, which could relax the so called Hubble tension: while direct measurements from Cepheid-calibrated type-Ia supernovae indicate \cite{Murakami:2023xuy}
\beq
H_0=\left( 73.29\pm 0.90 \right)\; {\rm km\over s\; Mpc}\,,
\eeq
CMB observables favor \cite{Planck:2018vyg}
\beq
H_0=\left( 67.71\pm 0.44 \right)\; {\rm km\over s\; Mpc}\,.
\eeq

In a recent work \cite{Cuesta:2021kca} we have proposed a variation of the $\Lambda$CDM model that is able to solve the $H_0$ tension by increasing $N_{\rm eff}$. 
As we briefly review in the next section, the key event in this new model is the resonant conversion 
of a fraction of CMB photons into an axion-like majoron (ALM) at the end of BBN.  Here we will analyze 
whether the sudden cooling of the universe at these or slightly higher temperatures implied by the conversion
may be consistent with the observed deuterium abundance, and also if it could have an impact on the long-standing lithium problem \cite{Fields:2011zzb}. A number of models have been proposed to solve this problem  (see discussion in \cite{ParticleDataGroup:2022pth}). Most of them imply the overabundance of deuterium  \cite{Erken:2011vv,Yamazaki:2014fja,Kusakabe:2014ola}, although new unstable MeV particles coupled to nucleons may reduce the Li abundance without affecting the one of lighter nuclei \cite{Goudelis:2015wpa}.

\section{Review of the ALM model}
The main particle physics motivation for the ALM are neutrino masses, as the model justifies their tiny value using approximate symmetries in an inverse seesaw mechanism \cite{Mohapatra:1986bd,Bernabeu:1987gr}. 
Let us be a bit more specific. We add 3 fermion singlets $(N,N^c,n)$ (we use 2-component spinors of left-handed chirality\footnote{In this 2-component notation $e_\alpha$ and ${e^c}_{\!\!\alpha}$ are the electron and the positron both {\it left}, whereas their conjugate-contravariant ${\bar e}^{\,\dot\alpha}$ and ${\bar {e^c}}\,^{\!\dot\alpha}$ are {\it right} spinors. The 4-component electron in the chiral representation of $\gamma^\mu$ is then 
$\Psi_e=\begin{pmatrix} e_\alpha \\ {\bar {e^c}}\,^{\!\dot\alpha} \end{pmatrix}$ [with $\bar \Psi_e=\left(  {e^c}\,^{\!\alpha}\; \bar e_{\dot \alpha} \right)$], while 
$\Psi_\chi=\begin{pmatrix} \chi_\alpha \\ {\bar {\chi}}\,^{\!\dot\alpha}\end{pmatrix}$ is a Majorana fermion.}) 
plus several scalar singlets $s_i$.
Then we impose a global $U(1)_X$ symmetry with the charge assignments in Table~\ref{table1}, and we assume an effective theory valid below a cutoff scale $\Lambda\approx 10$ TeV that suppresses higher dimensional operators. 
\begin{table}
\begin{center}
\begin{tabular}{|c|ccccccc|}
\hline
 & \hspace{0.1cm} $(\nu\; e)$ \hspace{0.1cm}& \hspace{0.1cm} $e^c$  \hspace{0.1cm} & 
 \hspace{0.1cm} $N$ \hspace{0.1cm}  & 
 \hspace{0.1cm} $N^c$  \hspace{0.1cm} & \hspace{0.1cm} $n$  \hspace{0.1cm} 
 & $(h^+\; h^0)$  & \hspace{0.1cm} $s_1,\, s_2,\, s_3,\, s_4\,\hdots$  \hspace{0.1cm} \\ [0.4ex]
\hline 
$Q_X$ & $+1$ &
$-1$ & $-2$ & $-1$ & $0$ & $0$ & $1,\,2,\, 3,\, 4,\,\hdots$ \\ [0.4ex]
\hline
$Z_3$ & $\alpha$ & 
$\alpha^*$ & $\alpha$ & $\alpha^*$ & $1$ & $1$ & $\alpha,\,\alpha^*\!,\,1,\,\alpha,\,\hdots$ \\ [0.4ex]
\hline
\end{tabular} 
\end{center}
\caption{Charges under the global symmetry and discrete $Z_3$ symmetry ($\alpha^3=1$).
\label{table1}}
\end{table}
Finally, we also assume that the scalar potential gives a VEV $v_X\approx $ TeV to a linear combination $s$ of scalar singlets that breaks the global symmetry and implies the presence of a (pseudo-)Goldstone boson $\phi$,  
\beq
s={1\over \sqrt{2}} \left( v_X + \rho \right) e^{i{\phi\over v_X}} \,.
\eeq
More precisely, we assume that $s$ is along the $s_3+\epsilon s_4$ 
direction in flavor space (the subindex indicates the $Q_X$ charge), with $\epsilon\ll 1$, and that the rest of scalar flavors are heavy (masses of order TeV). The charge $Q_X=3$ of the dominant component in the VEV defines then a $Z_3$ discrete symmetry, as $\langle s_3 \rangle$ is invariant under phase shifts $e^{-iQ_X\theta}$ with $\theta=2\pi/3$. This $Z_3$ symmetry forbids both standard neutrino masses $m_{\nu_i}$ and couplings $\lambda_{\nu_i}$ to the majoron. In turn, these terms appear with the {\it right} values
($m_{\nu_i}\le 0.1$~eV, $\lambda_{\nu_i}\approx m_{\nu_i}/m_X$) once the small component of the VEV along $s_4$ is included, breaking lepton number \cite{Cuesta:2021kca}. In addition, the possible presence of vectorlike charged leptons at the scale $\Lambda$ would induce
an axion-like coupling $g_{\phi\gamma\gamma}$ for the majoron. The result is a model with Majorana masses for the
standard neutrinos plus a light pseudoscalar with a Lagrangian
\beq
{\cal L} \supset i \, \lambda_{\nu_i}\,  \phi \; \overline \Psi_{\nu_i}  \gamma_5  \,\Psi_{\nu_i} - 
{1\over 4} \,g_{\phi\gamma\gamma}\, \phi \,\widetilde F_{\mu\nu} F^{\mu\nu}-
{1\over 2}\,m_\phi^2\;\phi^2\,,
\eeq
where we have assumed that gravitational effects break the global symmetry.
We take $m_\phi= 0.5$~eV, the majoron coupled mostly to the heaviest neutrino ($ \lambda_{\nu_3}=6.8\times 10^{-14}$) and a coupling to photons
$g_{\phi\gamma\gamma}=1.4\times10^{-11}\;{\rm GeV}^{-1}$. Such ALM would 
decay $\phi \to \nu \bar \nu$ with a lifetime
$\tau_\phi\approx 3.5\times 10^{12}$~s and a negligible branching ratio into photons.

The cosmological evolution of the model is specially interesting in the presence of a nG primordial magnetic field \cite{Subramanian:2015lua}: we assume a field with lines frozen in the plasma that evolves like $B\approx B_0 \,(T/T_0)^2$, with  $B_0=3$~nG and a coherence length $\lambda_0\ge 1$~Mpc. 
In the early universe photons get a plasmon mass and, due to the background magnetic field, they will mix with $\phi$. It turns out that as the universe cools it will cross a resonant temperature $\Tb$  where the ALM and the plasmon masses coincide and the oscillation probability of CMB photons into majorons can be easily estimated \cite{Ejlli:2013uda}. In the particular model described above \cite{Cuesta:2021kca}, 4.4\% of the photons carrying a 6.3\% of the CMB energy convert into ALMs at $\Tb=26$~keV. This implies a 1.6\% drop in the CMB temperature and, due to the sudden disappearance of photons, a 4.8\% increase in the baryon to photon ratio. The model avoids distortions in the CMB because {\it (i)}~at these temperatures photons are still in thermal equilibrium and {\it (ii)}~when finally the ALM decays it gives neutrinos, not gammas.

\begin{table}
\begin{center}
\begin{tabular}{|l|c|c|} 
\hline
  Parameter & $\Lambda$CDM & ALM \\ 
  \hline
$100\, \Omega_b h^2$           & $2.242\pm 0.014$ & $2.295\pm 0.014$\\ 
$\Omega_{\rm cdm} h^2$         & $0.119\pm 0.001$ & $0.129\pm 0.001$\\ 
$100\, \theta_s$               &$1.0420\pm 0.0003$& $1.0407\pm 0.0003$\\ 
$\ln \left( 10^{10}A_s\right)$ & $3.046\pm 0.015$ & $3.062\pm 0.016$ \\ 
$n_s$                          & $0.967\pm 0.004$ &  $0.991\pm 0.004$ \\ 
$\tau_{\rm reio}$              & $0.055\pm 0.008$ & $0.056\pm 0.008$\\ 
   \hline
$H_0$  [km/s/Mpc]              & $67.71\pm 0.44\;\,$  & $71.42\pm 0.50\;\,$ \\ 
\hline
\end{tabular}
\end{center}
\caption{Cosmological parameters in $\Lambda$CDM \cite{Planck:2018vyg} and the ALM model \cite{Cuesta:2021kca};
the baryon densities correspond to $\eta_{10}=6.14\pm 0.04$ and $\eta_{10}=6.28\pm 0.04$, respectively.
\label{table2}}
\end{table}

At $T<\Tb$ the universe evolves with a larger value of $\Omega_b$ and of $\Neff$ until $10 m_\phi\ge T\ge 0.1 m_\phi$. At these temperatures ({\it i.e.},  around recombination) decays and inverse decays ($\phi\leftrightarrow \nu \bar \nu$) become effective \cite{Escudero:2019gvw}, damping the free streaming of the extra radiation  and providing consistency with CMB observables. Once all majorons have decayed the final value of $\Neff$ is $3.85$. 
We have deduced the modified cosmological evolution of the model at these temperatures with the Boltzmann code {\tt CLASS} \cite{Blas:2011rf}, whereas the best fit of CMB \cite{Planck:2019nip} plus BAO \cite{BOSS:2016wmc,Ross:2014qpa,Beutler:2011hx} data is obtained with {\tt Monte Python} \cite{Audren:2012wb,Brinckmann:2018cvx}. In particular, we obtain the cosmological parameters given in Table \ref{table2}, where we also include our fit with $\Lambda$CDM for comparison. We find that 
the parameters fit the TTTEEE$\_$low$\ell\_$low$E\_$lensing dataset from Planck with a 
$\chi^2_{\rm ALM}=2784.9$, versus $\chi^2_{\Lambda{\rm CDM}}=2779.8$, yielding in both models a very similar goodness of fit (see Fig.~4 of \cite{Cuesta:2021kca}). Moreover, the fit of the BAO dataset (from BOSS, SDSS MGS, and 6dFGS) has
a $\chi^2_{\rm ALM}=5.3$, identical to the one that we obtain with $\Lambda$CDM.
Most important, 
the value of the Hubble constant in the ALM model is
\beq
H_0=\left( 71.42\pm 0.50 \right)\; {\rm km\over s\; Mpc}\,.
\eeq
This value would be consistent (within $1.8\sigma$) with the one from supernova observations calibrated with Cepheids,  versus a $5.6\sigma$ discrepancy in $\Lambda$CDM.
We wish to emphasize that although $\Delta\chi^2=\chi^2_{\rm ALM}-\chi^2_{\Lambda{\rm CDM}}=5.1$, the global fit including this prediction for $H_0$ gives $\Delta\chi^2=-22.6$, favouring the ALM model.

\section{BBN in the ALM model}

Let us discuss the implications of this ALM model for BBN. We will use the Mathematica code {\tt PRIMAT} \cite{Pitrou:2018cgg}\footnote{See \href{http://www2.iap.fr/users/pitrou/primat.htm}{http://www2.iap.fr/users/pitrou/primat.htm}.}, adapting the thermodynamics to the scenario under consideration. In particular, we assume that both the conversion of a fraction 
$r_\gamma=0.063$ of the CMB energy into majorons at $\Tb=26$~keV and the thermalization of the photons afterwards is instantaneous. 

The values of $\Omega_b h^2$ providing the best fit to CMB+BAO observables 
correspond to
\begin{eqnarray}
\eta_{10}&=6.14\pm 0.04\hspace{1cm} \Lambda\mbox{CDM} \\
\eta_{10}&=6.28\pm 0.04\hspace{1cm} \mbox{ALM} \,.
\end{eqnarray}
In the ALM model, 
however, this is the value at $T<\Tb$: at the beginning of BBN ({\it i.e.}, before the conversion of photons into majorons) the value is a factor of $(1-r_\gamma)^{3/4}$ smaller,  corresponding to $\eta_{10}=5.98$. 

\begin{figure*}
\includegraphics[scale=0.55]{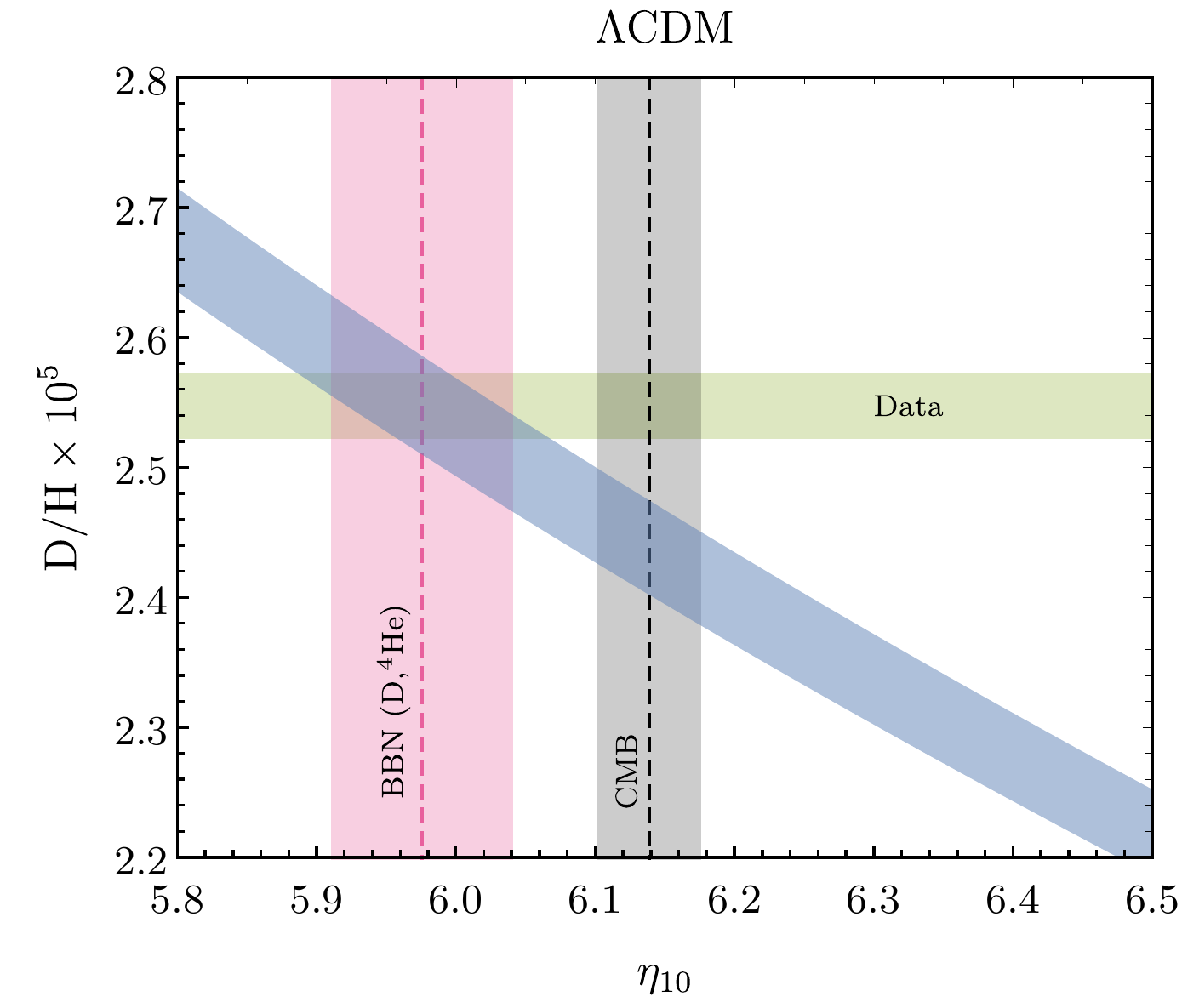}
\includegraphics[scale=0.55]{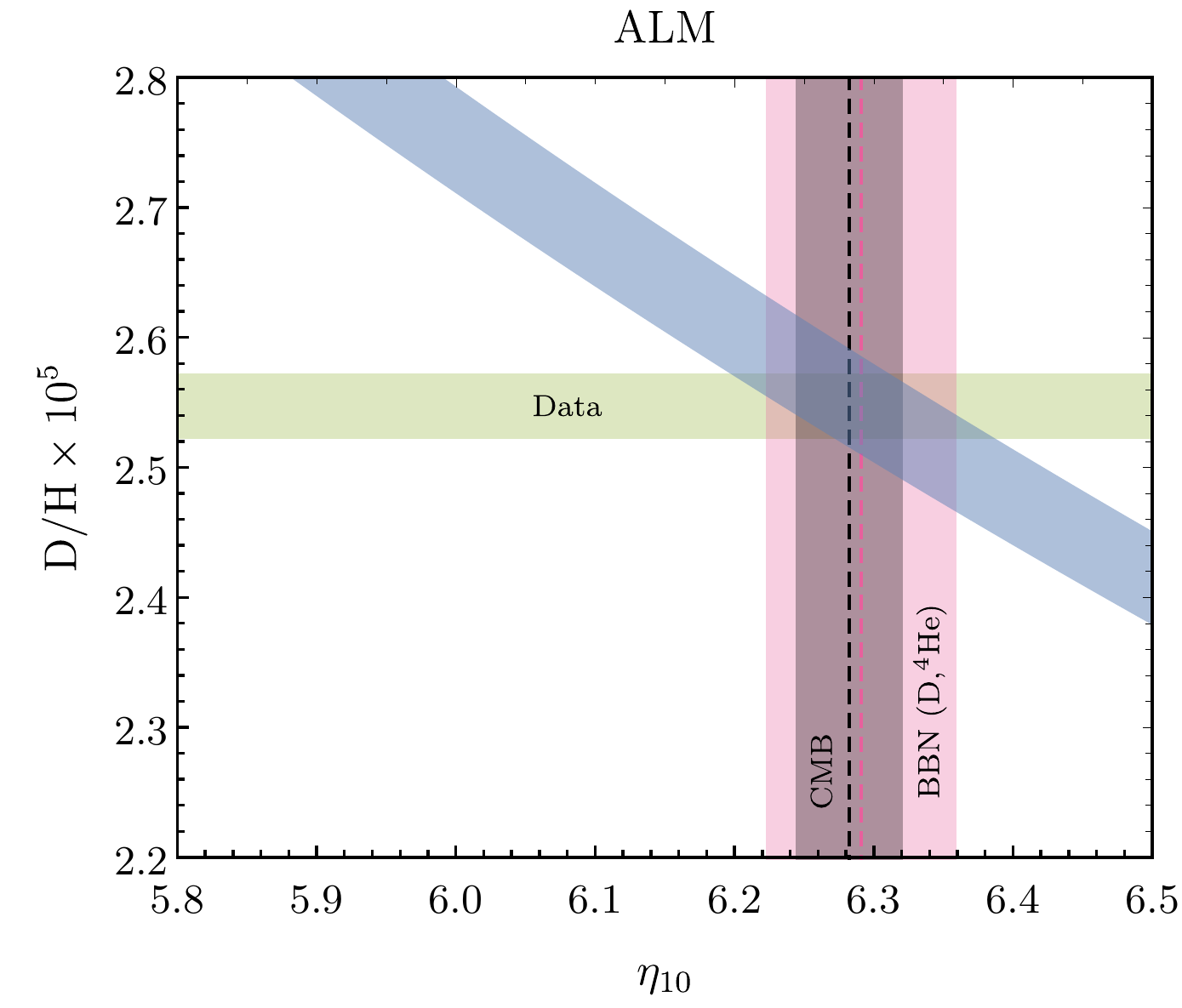}
\caption{Deuterium abundance for different values of $\eta_{10}$ in $\Lambda$CDM (left) and  
the ALM model (right).}
\label{fig1}
\end{figure*}

We can then compare the abundance of the primordial nuclei in the $\Lambda$CDM and ALM scenarios once the cosmological parameters have been fixed. Our results are summarized in Fig.~\ref{fig1}. The blue band expresses the number abundance of deuterium 
${\rm D/H}\equiv n_{\rm D}/n_{\rm H}$ for different values of $\eta_{10}$ in each model. The data (green band) gives \cite{ParticleDataGroup:2022pth}
\beq
{\rm D/H}=\left( 2.547 \pm 0.025 \right) \times 10^{-5}\,.
\eeq
Taking this abundance together with the observed $^4$He mass fraction  ${\rm Y_P}=\left( 0.245\pm 0.003\right)$ \cite{ParticleDataGroup:2022pth},
we obtain the value of $\eta_{10}$ preferred by BBN in each model (the pink band):\footnote{Notice that, incidentally, the value of 
$\eta_{10}$ preferred by BBN in $\Lambda$CDM coincides with the value of $\eta_{10}$ at $T>\Tb$ in the ALM model quoted above.}
\begin{eqnarray}
\eta_{10}&=5.98\pm 0.07\hspace{1cm} \Lambda\mbox{CDM} \\
\eta_{10}&=6.29\pm 0.06\hspace{1cm} \mbox{ALM} \,,
\end{eqnarray}
where the errors combine in quadrature the observational uncertainties in D/H and $^4$He with the theoretical uncertainty from nuclear rates (see Table VII in \cite{Pitrou:2018cgg}).
We can compare these values with the ones deduced from CMB+BAO observables in each model (the gray band); we see a slight tension (a 2$\sigma$ deviation, see also \cite{Pitrou:2020etk}) in $\Lambda$CDM versus a perfect agreement in the ALM scenario. We should mention, however, that other groups \cite{Pisanti:2020efz,Yeh:2020mgl} fit the observed deuterium abundance using a larger value of $\eta_{10}$, more in line with the values obtained from CMB data in $\Lambda$CDM. These groups base their analyses on 
a different energy dependence for the rates of ${\rm d\, d}\to {\rm n}\,^3{\rm He}$ and ${\rm d\, d}\to {\rm p}\,^3{\rm H}$ (see discussion in \cite{Pitrou:2021vqr}). The discrepancy underlines the need for precise measurements of these nuclear rates, that would minimize the impact of the energy dependence modeling. 

Notice that the $^7$Be in its neutral form decays via electron capture into $^7$Li with a lifetime of 77 days (about 1200 years in the primordial plasma \cite{Khatri:2010ed}).
The change in the lithium abundance obtained in the ALM model is small: 
$
({\rm ^7Li+^7\!Be})/{\rm H}= \left( 5.23 \pm 0.26 \right)\times 10^{-10}
$
versus $({\rm ^7Li+^7\!Be})/{\rm H}= \left( 5.50 \pm 0.25 \right)\times 10^{-10}$ in $\Lambda$CDM 
for the $\eta_{10}$ preferred by CMB observables in each model. Both values are far from the one required to explain the observed abundance \cite{ParticleDataGroup:2022pth},
\beq
{\rm ^7Li}/{\rm H} = \left( 1.6 \pm0.3 \right) \times 10^{-10}\,.
\eeq
Although the experimental determination of this abundance has been in permanent revision (it is unclear how the stars may have depleted it \cite{Fields:2022mpw}), we would like to generalize the scenario just described and see under what conditions it could explain an abundance of primordial lithium much smaller than the one obtained in $\Lambda$CDM.

\section{A generalized scenario}

There are two basic parameters that may impact BBN. The first one is the  temperature $\Tb$ for the resonant photon to axion conversion. In the ALM model discussed before we used $\Tb=26$~keV, a temperature where the synthesis of Li and Be has basically finished. Increasing $\Tb$ we may access an earlier stage where their production is still ongoing: the sudden cooling could interrupt it and imply a lower final abundance. The second parameter is the fraction $r_\gamma$ of CMB photon energy that is converted into ALMs. The larger $r_\gamma$ the stronger the cooling and thus the halt in the production of heavier nuclei.

The value of these two parameters is determined by the mass $m_\phi$ of the majoron, its coupling 
$g_{\phi\gamma\gamma}$ to photons and the strength of the background magnetic field. In addition, at $T<\Tb$ a given value of $r_\gamma$ implies a change in $\Neff$ from the high-temperature value\footnote{This value includes a small contribution $\Delta \Neff=0.026$ from thermal majorons decoupled at TeV temperatures \cite{Cuesta:2021kca}.}
 $\Neff^{\rm H}$ to 
\beq
\Neff^{\rm L}={\Neff^{\rm H}\over 1-r_\gamma} + {r_\gamma\over {7\over 8} \left(1-r_\gamma \right) \left( {4\over 11} \right)^{4/3}}
\eeq
and also a boost in the baryon to photon ratio, 
\beq
\eta^L_{10}={\eta^H_{10}\over (1-r_\gamma)^{3/4}}\,,
\eeq
with $\eta^H_{10}=5.99$ being the $\Lambda$CDM value when an extra $\Delta\Neff=0.026$ is included.
Large values of $\Neff$ do not necessarily conflict with  bounds from CMB and LSS observables, although their consistency would obviously require a more involved setup (if any). 
In particular, a large amount of non-interacting radiation at $T\lesssim \Tb$ ({\it e.g.}, $\Neff = 10$) may {\it not} free stream during recombination, and there are mechanisms to achieve that. First, it the majorons were heavier, {\it e.g.}, $m_\phi=10$ eV instead of $m_\phi=0.5$ eV, during recombination they would just appear as an extra contribution to the dark matter of the universe decaying later on into neutrinos.
Second, there could be {\it several} majorons of different mass: all it takes to reduce the neutrino free streaming during recombination is one of them with $m_\phi\approx 0.5$ eV, so that inverse decays ($\nu \bar \nu \to \phi$) become effective at this critical temperature. 
In any case, we think that it is interesting to explore how BBN observables may change when we let the two parameters $\Tb$ and $r_\gamma$ vary freely.

\begin{figure*}
\includegraphics[scale=0.55]{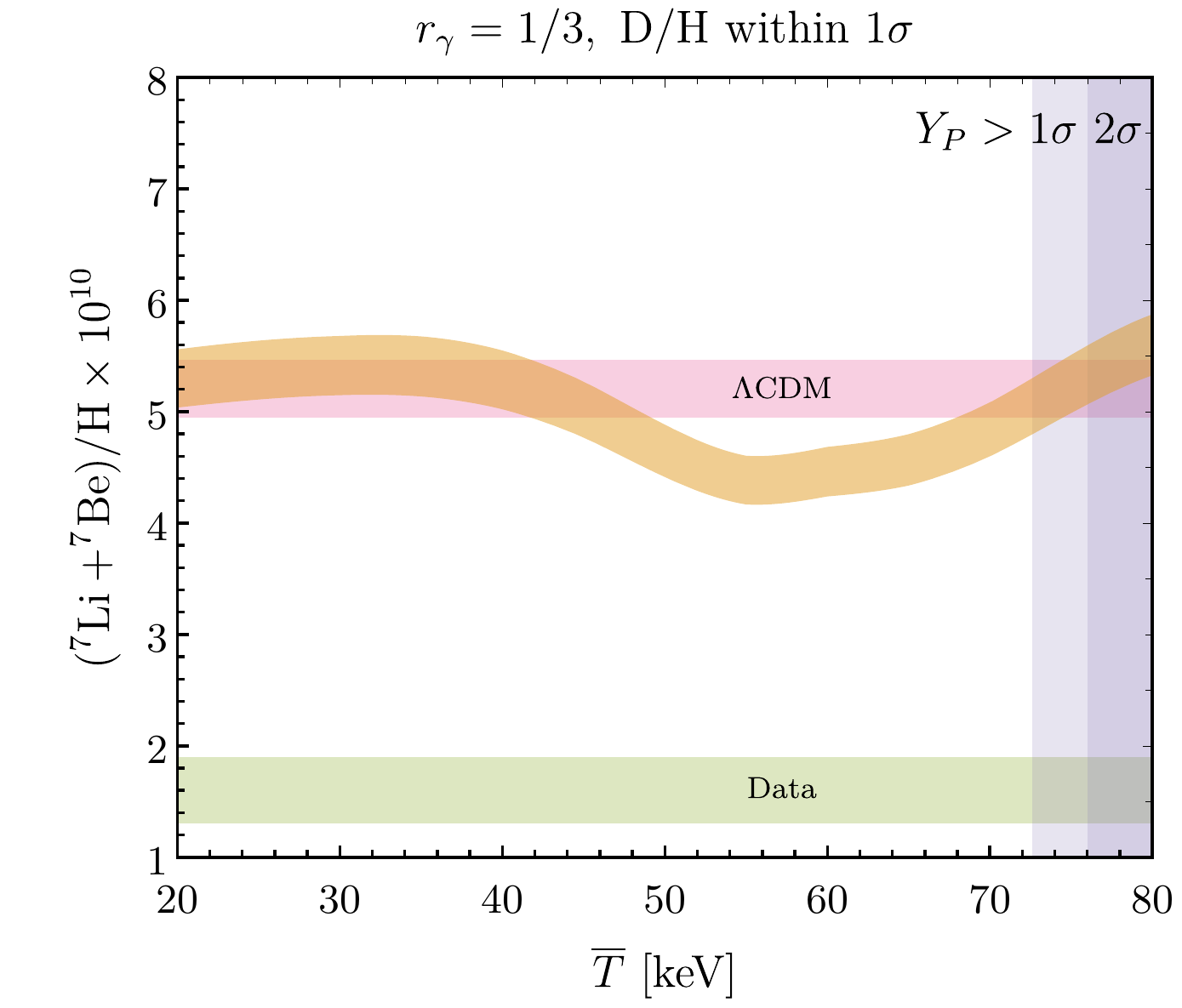}
\includegraphics[scale=0.55]{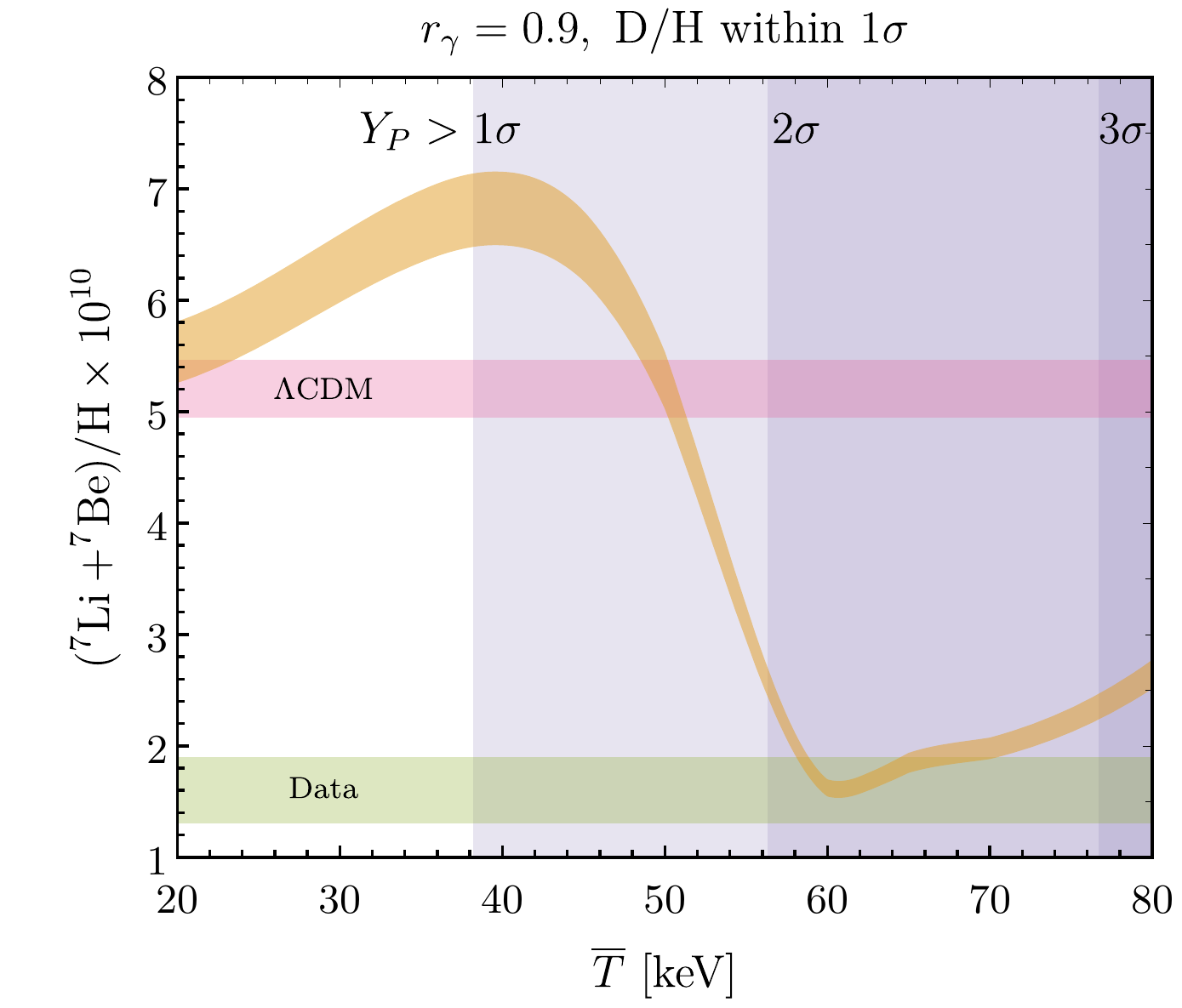}
\caption{Lithium abundance for different values of $\Tb$ and $r_\gamma=1/3$  (left) or $r_\gamma=0.9$ (right).}
\label{fig2}
\end{figure*}
Let us first consider the case with $r_\gamma=1/3$, corresponding to multiple oscillations and an effective chemical equilibrium between photons and majorons at $\Tb$. For each value of $\Tb$ between 20 and 80 keV, we find the value of 
$\eta_{10}$ that fits the deuterium abundance within 1$\sigma$. Then we obtain the abundance of $^4$He and $^7$Li+$^7$Be predicted by the model. We plot  our results in Fig.~\ref{fig2}-left. For $\Tb\approx 55$~keV, the lithium is reduced 
to $\left( 4.39 \pm 0.22 \right) \times 10^{-10}$ while the $^4$He abundance is still in agreement (within 1$\sigma$) with the data. This represents a 7.5$\sigma$ excess respect to the observed value, versus the 9.1$\sigma$ deviation obtained in $\Lambda$CDM 
when we use the value $\eta_{10}=5.98\pm 0.07$ that reproduces the deuterium abundance (the one preferred by CMB observables implies a $10\sigma$ excess).

\begin{figure}
\centering\includegraphics[scale=0.55]{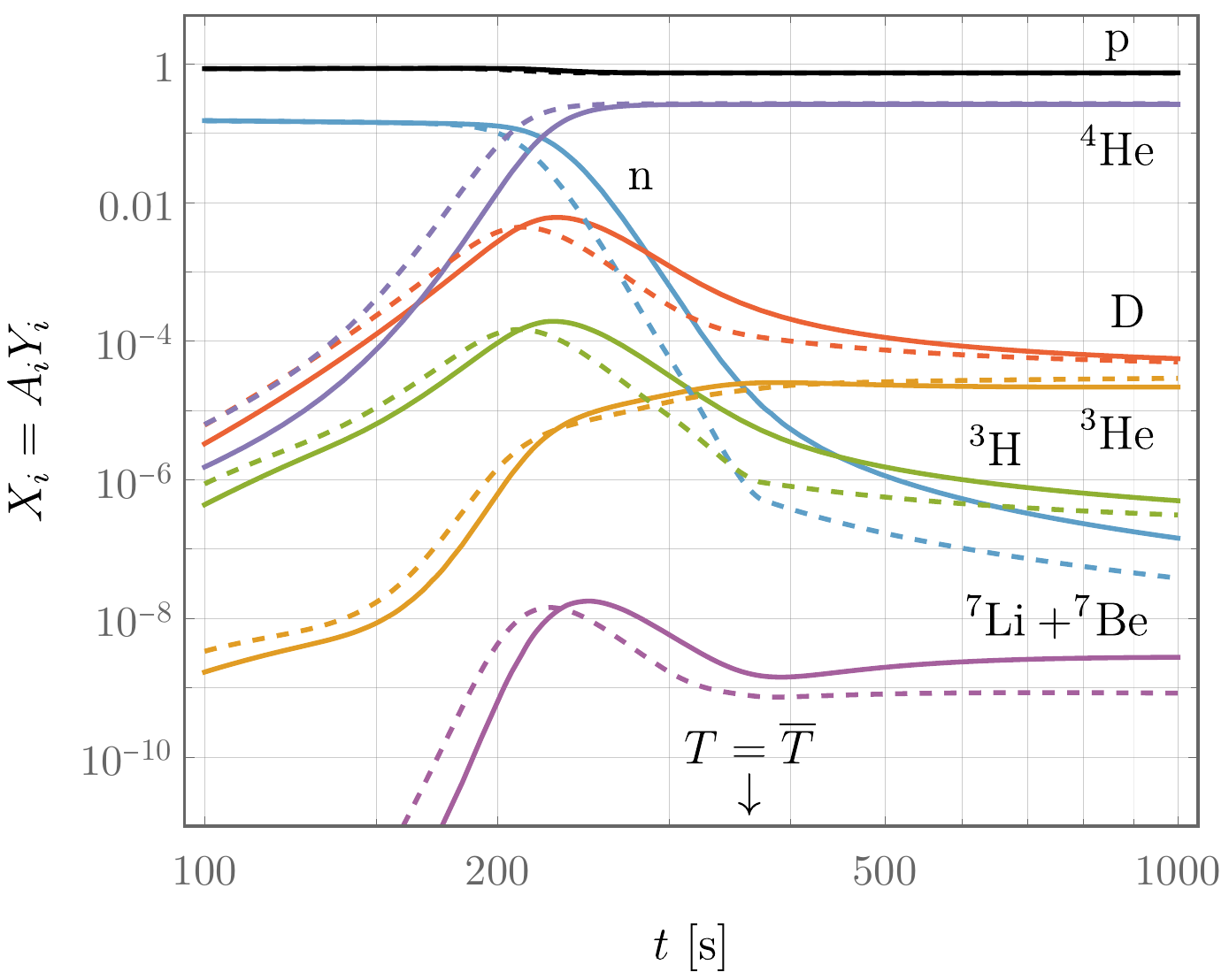} 
\caption{Primordial abundances at different times in $\Lambda$CDM (solid) and a model where $90$\% of the CMB energy converts into axions at $t\approx 360$ s ($\Tb=60$~keV) (dashes).}
\label{fig3}
\end{figure}

To obtain complete agreement we need a more extreme cooling: 90\% of the CMB energy converting into axions at $\Tb\approx 60$~keV (in Fig.~\ref{fig2}-right). More precisely, the deuterium and the lithium abundances predicted by the model coincide with the observed ones for $\Tb\approx 60$~keV, with just a 2$\sigma$ excess of $^4$He.
We can understand how this may happen studying the time evolution of the abundances in Fig.~\ref{fig3}. We start at higher temperatures with a much larger baryon to photon ratio: 
$\eta_{10}=11.44\pm0.12$ versus $5.98\pm 0.07$ in $\Lambda$CDM. This breaks the deuterium bottleneck at a higher temperature ($T\approx-B_{\rm D}/\ln\eta$, where $B_{\rm D}$ is the deuteron binding energy), and allows for an earlier synthesis of all nuclei. At this temperature the deuteron photodissociation, which prevents its fusion into other species, becomes less efficient.
Generically, the scenario implies an excess of the heavier $^4$He, $^3$He nuclei relative to the more reactive $n$, D, $^3$H, which allows for a better fit of the deuterium abundance (within one sigma of the experimental value for the chosen value of $\eta_{10}$). On the other hand, we see in the plot that
the sudden cooling at $\Tb$ {\it ends} BBN also earlier than in $\Lambda$CDM, freezing the $^7$Li+$^7$Be abundance when it was still growing. 
In particular, the reaction $^3{\rm He} \, ^4{\rm He}\to {^7{\rm Be}} \,\gamma$ is completely halted when it was becoming dominant, because the drop in temperature makes the reactives unable to overcome the Coulomb barrier. Furthermore, the reaction ${\rm p}\,^7{\rm Li}\to {}^4{\rm He}\,^4{\rm He}$ depletes lithium at the same time.
Overall, the result is a slight excess of $^4$He but a good global fit for all primordial nuclei.

\section{Summary and discussion}

The observation that the universe is made of 1 gram of $^4$He per 3 grams of H plus small amounts of other nuclei is considered one of the main successes of our cosmological model. During the past decade, however, the experimental precision achieved in observables related to CMB anisotropies, in the deuterium abundance (the data has reached 1\% precision and the uncertainties in the nuclear rates have been reduced \cite{Pisanti:2020efz}) 
or in the matter power spectrum test the model to a new level. Its consistency both with the data and with the standard model of particle physics is absolutely remarkable, which puts the focus on the few anomalies or {\it tensions} that can be identified. 

Here we have explored whether a model that was proposed to explain one of these tensions, the mismatch between the expansion rate of the universe now and during recombination, may also affect the two anomalies detected in BBN predictions, namely, {\it (i)}~the baryon to photon ratio preferred by the CMB takes the deuterium abundance 2$\sigma$ away from the data, and {\it (ii)}~the predicted lithium abundance is up to 10$\sigma$ larger than observed. These anomalies and also the $H_0$ tension are more of a call to revise our interpretation of the data than a strong reason to doubt the $\Lambda$CDM model. Still, we 
find it worth to consider possible variations of the model that may have an impact on these observables.

The ALM model is motivated by neutrino masses, as an inverse seesaw mechanism at the TeV scale requires the breaking of a global symmetry related to  lepton number and implies the presence of a light majoron decaying into neutrinos. The cosmological implications are also quite general:  a primordial magnetic field will mix any axion-like particle with the CMB photons, and as the universe expands it will scan a wide range of temperatures, eventually crossing the resonant temperature $\Tb$ for the photon to axion conversion. Indeed, we find somewhat surprising that the sudden cooling of the universe at the end of BBN through this mechanism has not been discussed before.

We have first considered the ALM model that relaxes the $H_0$ tension consistently with CMB observables. We have taken the cosmological parameters of the model (they were obtained with a customized version of {\tt CLASS} \cite{Blas:2011rf} and {\tt Monte Python} \cite{Audren:2012wb,Brinckmann:2018cvx}) and have used {\tt PRIMAT} \cite{Pitrou:2018cgg} to analyze its BBN predictions.
The model has a larger value of $\eta_{10}$ than $\Lambda$CDM during recombination but a smaller one during BBN. As a result, it {\it solves} the deuterium anomaly. 

Then we have studied under what conditions a generalized model with a possibly higher $\Tb$ and a larger fraction $r_\gamma$ of CMB energy converted into dark radiation may affect the lithium abundance. 
This possibility would probably require the presence of {\it several} ALMs, most of them heavier (as they should not free stream during recombination) and with weaker coupling to neutrinos (despite being heavier, they should also decay at $T<1$ eV). Of course, 
the motivation for such a bizarre scenario ({\it e.g.}, multiple pseudo-Goldstone bosons in a model with a larger global symmetry and a non-trivial pattern of breaking effects) would be weaker, but we find remarkable that it could imply a lithium abundance  3 times smaller than $\Lambda$CDM, as suggested by the data. 

The scenario discussed here would remind of the one proposed in \cite{Erken:2011vv}, where the CMB photons cool as they enter in thermal contact with an axion condensate that constitutes the dark matter of the universe. Also to the one with photon cooling by kinetic mixing with hidden photons considered in \cite{Jaeckel:2008fi}. In both cases the plasma temperature cools after BBN,  implying an overproduction of deuterium. What makes our case {\it different} is {\it (i)}~that the conversion is resonant, at a precise and common temperature for CMB photons of all energies (although with a different probability \cite{Cuesta:2021kca}) {\it (ii)}~that it occurs at a high temperature, so that the CMB photons rethermalize erasing all spectral distortions and {\it (iii)}~that the model may incorporate a mechanism to reduce the free streaming of the dark radiation during recombination (ALM decays and inverse decays \cite{Escudero:2019gvw}).

None of the current tensions in cosmology have won general acceptance: observations must still fully disentangle 
BBN from effects like the depletion and synthesis of nuclei in stars or their production through cosmic ray
spallation, while the Hubble tension should be seen within a complete and calibration-independent picture. 
In the near future, however, the quality and the 
quantity of the probes will only improve, eventually telling apart $\Lambda$CDM from variations like the one discussed here.

\acknowledgments

We would like to thank Carlos Abia for useful discussions.
This work has been partially supported by the Spanish Ministry of Science, Innovation and Universities (PID2019-107844GB-C21/AEI/10.13039/501100011033) and by Junta de Andaluc{\'\i}a/FEDER (P18-FR-5057). 
It is also based based upon work from the COST Action COSMIC WISPers CA21106,
supported by COST (European Cooperation in Science and Technology).
AJC acknowledges funding from the European Union - NextGenerationEU and the Ministerio de Universidades of Spain through {\it Plan de Recuperaci\'on, Transformaci\'on y Resiliencia}. 

%\paragraph{Note added.} This is also a good position for notes added after the paper has been written.

%\bibliography{paper.bib}{}
%\bibliographystyle{JHEP}

\providecommand{\href}[2]{#2}\begingroup\raggedright\endgroup

\end{document}